\documentclass[a4paper, sort&compress]{elsarticle}
\usepackage{ucs}
\usepackage[utf8]{inputenc}
\usepackage{amsmath,amssymb,amsfonts,mathtools}
\usepackage[english]{babel}
\usepackage{fontenc}
\usepackage{graphicx}
\usepackage{ulem}
\usepackage[margin=0pt,font=footnotesize, labelsep=colon]{caption}
\usepackage{tabularx}
\usepackage{multirow}
\usepackage{units}
\usepackage{booktabs}
\usepackage{bbm}
\usepackage{eurosym}
\usepackage{color}
\usepackage[table,dvipsnames]{xcolor}
\usepackage{colortbl}
\usepackage{supertabular}
\usepackage{array,lipsum}
\usepackage{lineno}
\usepackage[hidelinks]{hyperref}

\definecolor{rgrey}{gray}{0.75}
\newcommand{\mean}[1]{\langle #1 \rangle}

\newcommand{\fref}[1]{Figure~\ref{#1}}
\newcommand{\sref}[1]{Section~\ref{#1}}
\newcommand{\tref}[1]{Table~\ref{#1}}

\usepackage[noabbrev]{cleveref}
\usepackage[export]{adjustbox}[2011/08/13]

\usepackage{algorithm}
\usepackage[noend]{algpseudocode}
\usepackage{placeins}
\usepackage{subcaption}
\usepackage[toc,nonumberlist]{glossaries}
\usepackage{float}
\newfloat{algorithm}{H}{lop}
\floatname{algorithm}{Algorithm}

\newacronym{vres}{VRES}{Variable Renewable Energy Sources}
\newacronym{rea}{REA}{Renewable Energy Atlas}
\newacronym{ncep}{NCEP}{American National Centers for Environmental Prediction}
\newacronym{cfsr}{CFSR}{Climate Forecast System Reanalysis}
\newacronym{ac}{AC}{Alternating Current}
\newacronym{dc}{DC}{Direct Current}
\newacronym{lcoe}{LCOE}{Levelised Cost of Electricity}
\newacronym{hvdc}{HVDC}{High Voltage Direct Current}
\newacronym{iset}{IWES}{Fraunhofer-Institut f\"ur Windenergie und Energiesystemtechnik}
\newacronym{ga}{GA}{Genetic Algorithm}
\newacronym{gas}{GAS}{Greedy Axial Search}
\newacronym{cs}{CS}{Cuckoo Search}
\newacronym{de}{DE}{Differential Evolution}
\newacronym{CapEx}{CapEx}{capital expenditures}
\newacronym{OpEx}{OpEx}{operational expenditures}

\usepackage{tikz}
\usepackage{tkz-graph}
\usetikzlibrary{backgrounds}
\GraphInit[vstyle = Normal] 
\tikzset{
  VertexStyle/.append style = { inner sep=0pt,minimum size=18pt,
                                font = \normalsize},
  EdgeStyle/.append style = {->,>=triangle 45} 
}
\usetikzlibrary{shapes.geometric}
\usetikzlibrary{shapes.arrows}

\definecolor{AC}{rgb}{0.7, 0, 0}
\definecolor{DC}{rgb}{0, 0.8, 0}

\begin{document}
\crefname{appendix}{}{}

\begin{frontmatter}

\title{
Flow-based nodal cost allocation in a heterogeneous highly renewable European electricity network
}
\author[label1,label2]{Bo Tranberg}
\ead{bo@eng.au.dk}
\author[label3]{Leon J. Schwenk-Nebbe\fnref{fn1}}
\ead{leon@schwenk-nebbe.com}
\author[label1]{Mirko Sch\"{a}fer}
\ead{schaefer@eng.au.dk}
\author[label4]{Jonas H\"{o}rsch}
\ead{hoersch@fias.uni-frankfurt.de}
\author[label1]{Martin Greiner}
\ead{greiner@eng.au.dk}
\fntext[fn1]{Present address: {\O}rsted, Teknikerbyen 25, 2830 Virum, Denmark}
\address[label1]{Department of Engineering, Aarhus University, Inge Lehmanns Gade 10, 8000 Aarhus C,  Denmark}
\address[label2]{Danske Commodities A/S, Vaerkmestergade 3, 8000 Aarhus C, Denmark}
\address[label3]{Department of Physics and Astronomy, Aarhus University, Ny Munkegade 120, 8000 Aarhus C,  Denmark}
\address[label4]{Frankfurt Institute for Advanced Studies (FIAS), Johann Wolfgang Goethe Universit\"at, Ruth-Moufang-Stra{\ss}e 1, 60438 Frankfurt am Main, Germany}

\begin{abstract}
For a cost efficient design of a future renewable European electricity system, the placement of renewable generation capacity will seek to exploit locations with good resource quality, that is for instance onshore wind in countries bordering the North Sea and solar PV in South European countries. Regions with less favorable renewable generation conditions benefit from this remote capacity by importing the respective electricity as power flows through the transmission grid. The resulting intricate pattern of imports and exports represents a challenge for the analysis of system costs on the level of individual countries. Using a tracing technique, we introduce flow-based nodal levelized costs of electricity (LCOE) which allow to incorporate capital and operational costs associated with the usage of generation capacity located outside the respective country under consideration. This concept and a complementary allocation of transmission infrastructure costs is applied to a simplified model of an interconnected highly renewable European electricity system. We observe that cooperation  between the European countries in a heterogeneous system layout does not only reduce the system-wide LCOE, but also the flow-based nodal LCOEs for every country individually.
\end{abstract}

\end{frontmatter}

\section{Introduction}
\label{sec:introduction}
A future sustainable electricity system will strongly depend on the efficient integration of high shares of renewable power generation (see for instance the Energy Roadmap 2050 from the European Commission~\cite{roadmap2050}). In particular, wind and solar technologies have become increasingly cost competitive~\cite{IRENA2018} and show considerable expansion potential for large-scale deployment~\cite{EEA2009,Suri2007}. The weather-dependent resource quality and thus cost efficiency of these variable renewable energy sources (VRES) is unevenly distributed across the European continent. An efficient placement of generation capacity will result in a heterogeneous layout, in which locations with favorable conditions will be net exporters of electricity, whereas regions with less favorable conditions import electricity as power flows through the transmission grid. In~\cite{eriksen2017} this cost benefit of a heterogeneous system layout has been shown for a simplified model of a highly renewable European electricity system. The influence of different levels of transmission capacity expansion for the total cost and the system layout has been studied in~\cite{schlachtberger2017}. Despite its efficiency in terms of reducing global system costs, such a heterogeneous layout represents a political and economic challenge. Countries with favorable weather conditions will get assigned disproportionally high shares of generation capacity, which to a large degree will be exported and serve electricity consumption abroad. The associated investment costs thus must be incentivized by appropriate remuneration schemes, which in today's system are largely based on electricity markets complemented with different kinds of state-regulated support schemes (see~\cite{stoft} for a general introduction into Power System Economics). Nevertheless, as discussed in~\cite{Morales2013}, the increasing share of fluctuating renewable generation represents a challenge for the design of future market rules. A deeper understanding of the nodal structure of system costs might provide guidance to the development of a suitable regulatory framework, which supports the transition towards an efficient sustainable system design by providing a fair allocation  of costs and target-oriented incentives for investors.

Using the flow tracing technique based on average participation introduced in~\cite{Bialek1996} and~\cite{ Kirschen1997}, we derive nodal levelized costs of electricity (LCOE) which factor in the share of the system-wide operational and capital costs associated with the electricity consumption (load) of the individual nodes. We apply this formalism to a coarse-grained model of the European electricity system with a high share of renewable generation, which has been studied in~\cite{eriksen2017}.

This paper is organized as follows: \sref{sec:model} describes the simplified model of the European electricity network and presents the respective infrastructure measures, cost modelling and heterogeneous renewable capacity layouts. Furthermore, the method of flow tracing is reviewed, which represents the cornerstone of the flow-based nodal cost allocation defined in~\sref{sec:nodal_cost_allocation}. In~\sref{sec:results} the application of this method is discussed for different renewable generation layouts and transmission cost allocation schemes. The paper concludes with a discussion of the results and an outlook to future research.

\section{Modelling and Methods}
\label{sec:model}
\subsection{The electricity network}
\begin{figure}[h!] \centering
\includegraphics[width = \columnwidth]{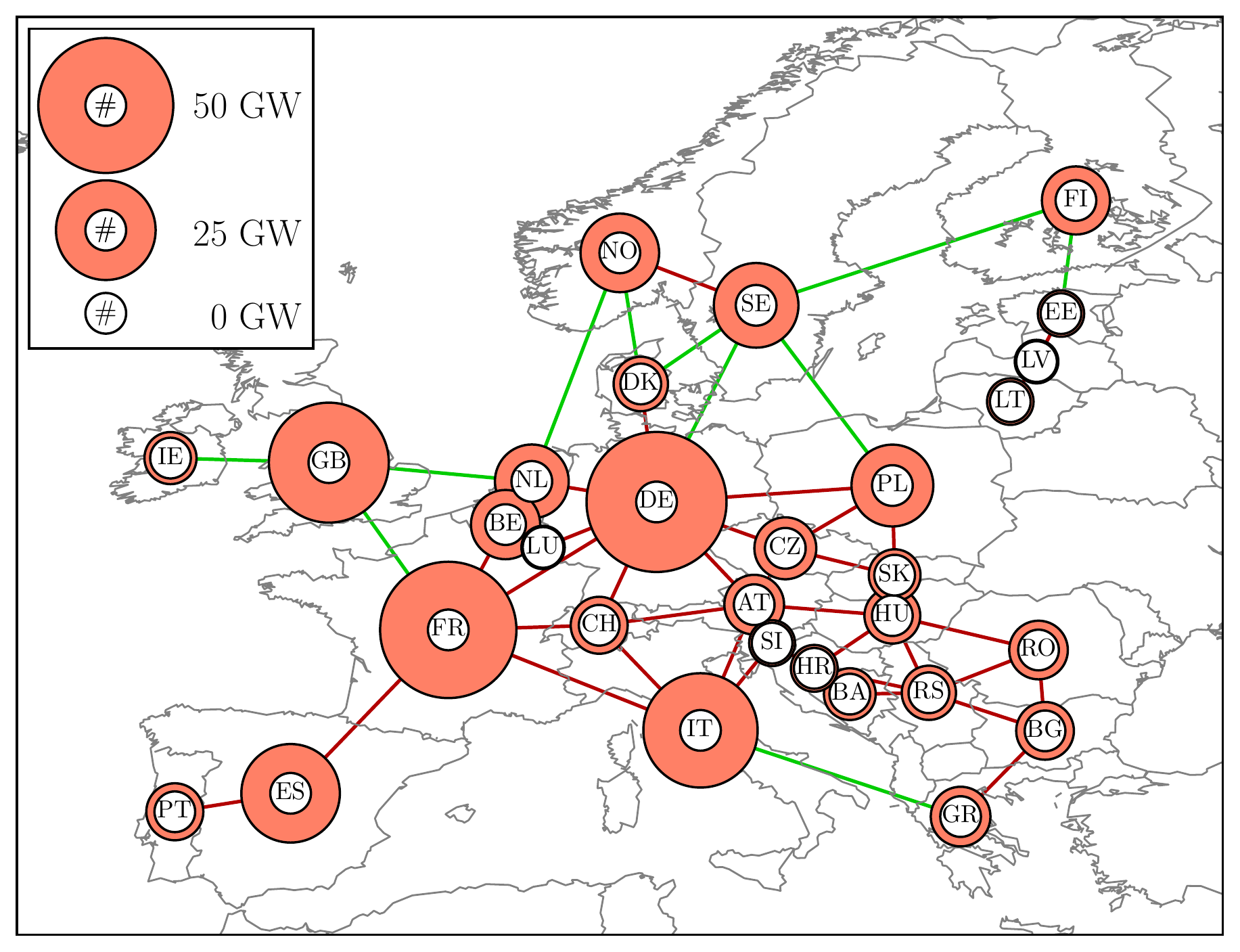}
\caption{The simplified European electricity system. The size of the nodes shows the average load of the respective countries. AC lines are colored red and DC lines are colored green.}
\label{fig:map}
\end{figure}

In this study we use a simplified model of the European electricity network, shown in~\fref{fig:map}. Each node represents an aggregated country and each link represents the coarse-grained interconnector transmission capacity between the countries, distinguishing between AC lines and HVDC lines. The size of the nodes visualizes the average load.
Wind and solar PV generation constitute the nodal VRES power generation:
\begin{equation}\label{eq:generation}
G_{n}^{R}(t) = G_{n}^{W}(t) + G_{n}^{S}(t)~.
\end{equation}
As described in~\cite{eriksen2017} the VRES generation is modeled using eight years of hourly weather data from 2000 to 2007 with a spatial resolution of $50\times 50 \text{km}^2$~(see~\cite{bofinger} and~\cite{heide2010} for details). The weather data is converted into generation time series using country-specific capacity layouts from Eurostat~\cite{eurostat1, eurostat2}. For simplicity this model focuses only on onshore wind and solar PV generation. We use two parameters to describe the renewable generation in \eqref{eq:generation}. The renewable penetration, which determines the amount of renewable generation relative to the load of a node
\begin{equation}
\langle G_{n}^{R} \rangle = \gamma_{n}\langle L_{n} \rangle ~,
\end{equation}
where the load time series $L_{n}(t)$ is based on historical data from ENTSO-E, and the mixing parameter $\alpha_n$, which fixes the ratio between wind and solar generation
\begin{align}
\langle G_{n}^{W} \rangle &=\alpha_{n} \langle G_{n}^{R} \rangle~,\\
\langle G_{n}^{S} \rangle &= (1-\alpha_{n})\langle G_{n}^{R} \rangle~.
\end{align}
Note the usage of the mean which implies that $\gamma_n=1$ describes a country that on average covers its entire load by renewable generation. The hourly nodal mismatch between VRES generation and load
\begin{equation}
\Delta_{n}(t) = G_{n}^{R}(t) - L_{n}(t)
\end{equation}
will be balanced by the dispatch of backup power generation, curtailment of excess generation, and power transmission between the countries. For simplicity, storage is not considered. The resulting nodal balancing equation reads
\begin{equation}
G_{n}^{R}(t) - L_{n}(t) =
\Delta_{n}(t) =
B_{n}(t) + P_{n}(t)~.
\label{eq:balancing}
\end{equation}
Here $B_n(t)$ represents the nodal balancing consisting of curtailment of excess power $C_n(t)=\max(B_n(t),0)$ and the dispatch of backup generation $G^B_n(t)=-\min(B_n(t),0)$, and $P_n(t)$ represents the power injected (ejected) to (from) the network. It is assumed that the model is balanced such that $\sum_{n}P_{n}(t)=0$. We assumed that the dispatchable backup generation is realized by Combined Cycle Gas Turbines (CCGT). The dispatch of the nodal backup and curtailment is determined using the synchronized balancing scheme (see~\cite{rodriguez2015localized} for a discussion of this dispatch mechanism):
\begin{equation}
\label{eq:synchronized_balancing}
B_{n}(t) = \frac{\langle L_{n} \rangle}{\sum_{k}\langle L_{k} \rangle}
\sum_{m}\Delta_{m}(t)~.
\end{equation}
Combining \eqref{eq:balancing} and \eqref{eq:synchronized_balancing} fixes the injection pattern~$P_{n}(t)$. The injection pattern in turn determines the flows~$F_{l}(t)$ on the links~$l$:
\begin{equation}
F_{l}(t) = \sum_{n}H_{ln}P_{n}(t)~.
\end{equation}
Here we have used the DC approximation to the AC power flow equations, in which the entries $H_{ln}$ of the matrix $\mathbf{H}$ are the power transfer distribution factors (PTDF) representing the influence of the line susceptances and network topology on the power flows. The usefulness of this approximation for active power flow analysis has been discussed in~~\cite{purchala2005}. The PTDF matrix~$\mathbf{H}$ can be calculated as
\begin{equation}
\mathbf{H} = \mathbf{\Omega}\mathbf{K}^{T}\mathbf{B}^{\dagger}~,
\end{equation}
where $\mathbf{B}^{\dagger}$ denotes the Moore-Penrose pseudo inverse of the nodal susceptance matrix $\mathbf{B}$ and the entries of the diagonal matrix $\mathbf{\Omega}$ are the line susceptances on the links $l$. The matrix $\mathbf{K}^{T}$ is the transposed incidence matrix with
\begin{equation}
K^{T}_{ln}= \left\{%
\begin{array}{ll}
1 & \mbox{if link $l$ starts at node $n$}~,\\
-1 & \mbox{if link $l$ ends at node $n$}~,\\
0 & \mbox{otherwise}~.
\end{array}\right.
\end{equation}
For simplicity the susceptances are all chosen to be identical and $\mathbf{\Omega}$ is the identity matrix.
\subsection{Infrastructure measures}
The energy system cost is based on measures of the installed generation and transmission capacities as well as on the backup energy introduced in~\cite{rodriguez2015cost}. The nodal backup energy is given by the average backup power generation:
\begin{equation}
E^B_{n} = \mean{G_n^B}~.
\end{equation}
Excluding extreme events that are assumed to be covered by emergency equipment or flexible demand outside the model, the backup capacity is defined as the $99\%$ quantile of the backup generation events:
\begin{equation}
0.99 = \int_{0}^{\mathcal{K}_{n}^{B}} d G_{n}^{B}p_{n}(G_{n}^{B})~.
\end{equation}
The transmission capacity $\mathcal{K}_{l}^{T}$ is defined in a similar way as the $99\%$ quantile of flow events,
\begin{equation}
\label{eq:transmission_capacity}
0.99 = \int_{-\mathcal{K}_{l}^{T}}^{\mathcal{K}_{l}^{T}} dF_{l} p_{l}(F_{l})~,
\end{equation}
assuming identical capacity in both directions. The total backup capacity is calculated by summing the nodal capacities $\mathcal{K}^{B}=\sum_{n}\mathcal{K}_{n}^{B}$. The total transmission capacity is calculated as the weighted sum
\begin{equation}
\mathcal{K}^{T} = \sum_{l}d_{l}\mathcal{K}_{l}^{T}~,
\end{equation}
taking into account the link length $d_l$ approximated by the distance between the capitals of the countries. From the renewable penetration and the wind/solar mix we derive the capacities of wind and solar generation:
\begin{align}
\mathcal{K}_{n}^{W} &= \frac{\gamma_{n}\alpha_{n}\langle L_{n} \rangle}{\text{CF}_{n}^{W}}~,\\
\mathcal{K}_{n}^{S} &= \frac{\gamma_{n}(1-\alpha_{n})\langle L_{n} \rangle}{\text{CF}_{n}^{S}}~.
\end{align}
The capacity factors $\text{CF}_{n}^{W/S}$ represent the average renewable generation as a fraction of the installed capacity. They are taken from~\cite{eriksen2017} where they are based on data from~\cite{eurostat1} and~\cite{eurostat2}.

\subsection{Cost modelling}
We follow the cost modelling as presented in~\cite{eriksen2017}. The present value of investment $V$ for each type of generation capacity is defined as
\begin{equation}
\label{eq:present_value}
V = \text{CapEx} + \sum_{t=1}^{T_{\text{life}}}\frac{\text{OpEx}_{t}}{(1+r)^{t}}~,
\end{equation}
where $r$, the rate of return, is assumed to be $4\%$ per year. Cost assumptions for capital expenditures (CapEx) and operational expenditures (OpEx) are listed in \tref{tab:costs}.
\begin{table}[!t]
  \centering
  \caption{Cost assumptions separated into capital expenditures (CapEx) and fixed and variable operational expenditures (OpEx) as well as expected lifetimes.}
      \label{tab:costs}
  \begin{tabular}{l|rrrr}
  \toprule
    \textbf{Asset} & \textbf{CapEx} & \textbf{OpEx$_{\text{fixed}}$} & \textbf{OpEx$_{\text{variable}}$} & \textbf{Lifetime}\\
    & [\euro/W] & [\euro/kW/y] & [\euro/MWh] & [years]\\
    \midrule
    CCGT & 0.90 & 4.5 & 56.0 & 30\\
    Solar PV & 0.75 & 8.5 & 0.0 & 25\\
    Onshore wind & 1.00 & 15.0 & 0.0 & 25\\
    \bottomrule
  \end{tabular}
\end{table}

The present value of a transmission line is calculated as the cost of the line
\begin{equation}\label{eq:trans-cost}
V_l^T = \mathcal{K}_l^Td_l c_l,
\end{equation}
where $d_l$ is the line length and $c_l$ is the specific transmission capacity cost:
\begin{equation}
c_l = \begin{cases}
400\text{\euro/km} & \text{(AC line),} \\ 1500\text{\euro/km} & \text{(DC line).}
\end{cases}
\end{equation}
The total present value of the transmission system is
\begin{equation}\label{eq:trans-value}
V^T = \sum_l V^T_l + N_\text{HVDC}\cdot 150,000 \text{\euro}.
\end{equation}
The second term accounts for the cost of a pair of converter stations for each HVDC line~(see \cite{eriksen2017} for all cost assumptions). The layout of AC and HVDC lines has been constructed in~\cite{rodriguez2014transmission} based on the existing European network in the year 2011 as described in~\cite{ENTSOEcaps}. New predicted lines until 2014 between Norway and the Netherlands~\cite{NorNed} and between Great Britain and the Netherlands~\cite{BritNed} have been added to this topology.

The system costs are measured using the Levelized Cost of Electricity (LCOE), which represents the average cost per consumed unit of energy during the system lifetime (see~\cite{kost2012levelized} for a discussion of the LCOE of renewable energy technologies):
\begin{equation}
\label{eq:lcoe}
\text{LCOE}_{V} = \frac{V}{
\sum_{t=1}^{T_{\text{life}}}\frac{L_{EU,t}}{(1+r)^{t}}
}~.
\end{equation}
The system $\text{LCOE}$ is calculated by summation over the $\text{LCOE}$s for the system elements, which takes into account the different life times for solar PV, onshore wind, CCGT plants, and transmission infrastructure. For $V\in\{V_W,\ W_S,\ V_B,\ V_T\}$:
\begin{equation}
    \label{eq:system-lcoe}
    \text{LCOE}_\text{EU} = \sum_V \text{LCOE}_V\,.
\end{equation}

\subsection{Optimal heterogeneity}
All countries have a natural upper limit of the geographical potential for renewable generation capacity. A lower limit is set by the willingnesses of the countries to be dependent on imports from other countries. A stylized model of heterogeneity within these boundaries, which avoids specifying the exact renewable potentials in each country, is introduced by the heterogeneity parameter $K$:
\begin{equation}\label{eq:heterogeneity-constraint}
\frac{1}{K}\leq \gamma_{n} \leq K~.
\end{equation}
The specific choice $K=1$ results in a homogeneous layout, in which the average load $\langle L_{n}\rangle$ corresponds to the average renewable power generation $\langle G_{n}^{R}\rangle$ in each country. In~\cite{eriksen2017} this condition was used to optimize the set of $\gamma_n$ and $\alpha_n$, a total of 60 variables, with the objective to minimize the system LCOE, and with the constraint that the renewable penetration of Europe is $\gamma_\text{EU}=1$. This was done using a Greedy Axial Search (GAS) algorithm~(see~\cite{cormen2001introduction} for a description). The optimized layouts are refered to as GAS layouts. In the following, GAS $\to$ GAS* refers to an additional optimization, in which the transmission capacities have been uniformly scaled down from the definition in \eqref{eq:transmission_capacity}, to yield a more cost effective constrained system despite a slightly higher cost for backup energy. This procedure has been discussed in~\cite{eriksen2017}.

\begin{figure}[h!] \centering
\includegraphics[width = \columnwidth]{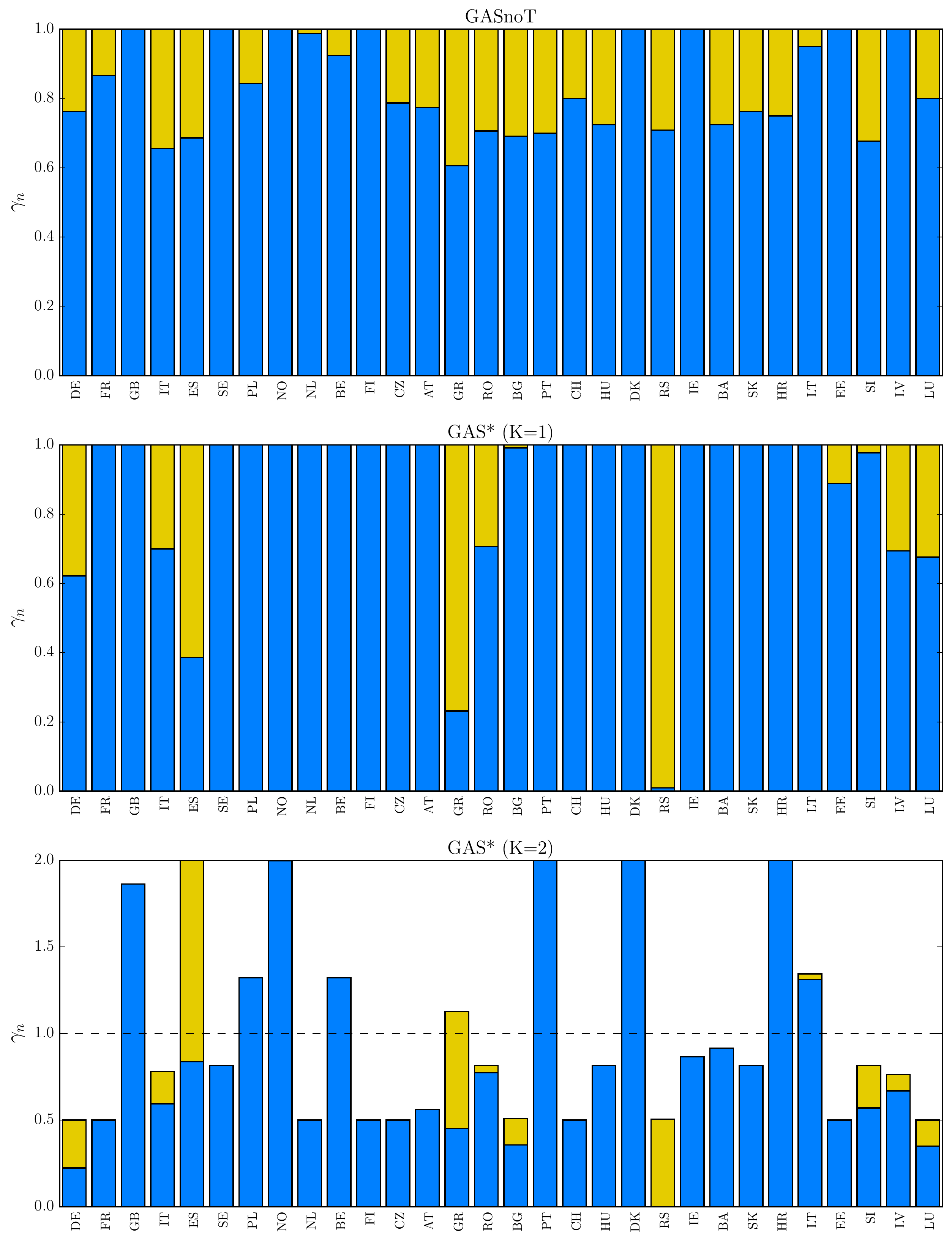}
\caption{The renewable penetration parameter $\gamma_n$ for GASnoT, GAS* K=1, and GAS* K=2 obtained in \cite{eriksen2017}. The mixing parameters $\alpha_n$ between wind (blue) and solar (yellow) power generation are also indicated.}
\label{fig:gamma_layout}
\end{figure}

\fref{fig:gamma_layout} shows the sets $\gamma_n$ and $\alpha_n$ of three optimized scenarios. The GASnoT layout in the top panel assumes a European system without transmission capacity between the individual countries. Setting the parameter $\gamma_{n}=1$ for all nodes $n$, we assume that on average every country individually covers its load from renewable generation, with the instantaneous mismatch balanced by local backup power generation. This leaves the nodal mix between wind and solar $\alpha_{n}$ to be optimized by the GAS algorithm. For countries with strong wind resources like Denmark (DK) or Sweden (SE), a $100\%$ wind layout is optimal, whereas southern countries like Italy (IT) or Greece (GR) introduce higher shares of solar power into the renewable energy mix. The middle panel represents a system allowing transmission between the nodes, allowing a more efficient placement of generation capacity and providing a spatial smoothing of fluctuations in the renewable generation. Whereas the GAS* (K=1) case still assumes a homogeneous parameter $\gamma_{n}=1$ for all nodes, the optimized mix of renewable generation represented by the parameters $\alpha_{n}$ differs from the GASnoT scenario due to the possibility of power transmission. The GAS* (K=2) scenario describes a heterogenous system layout, in which also the parameters $\gamma_{n}$ have been optimized under the condition \eqref{eq:heterogeneity-constraint} with $K=2$ and an overall VRES penetration of $\gamma_{EU}=1$. The bottom part of \fref{fig:gamma_layout} illustrates this heterogeneity: solar and wind generation capacity is concentrated at locations with favorable resource quality, leading to higher VRES penetration $\gamma_{n}$ for these nodes. In this layout, nodes with $\gamma_{n}>1$ are net exporters, whereas nodes with $\gamma_{n}<1$ are net importers.

The overall system LCOE for the three layouts of \fref{fig:gamma_layout} are as follows. The GASnoT layout without transmission has the highest cost of~63~\euro/MWh. The more efficient placement of renewable capacity in the GAS* K=1 layout leads to a reduction of system LCOE to~56.6~\euro/MWh, just by introducing transmission. When introducing additional heterogeneity in the GAS* K=2 layout the resulting power flows in the system lead to higher transmission capacity costs. However, the total system LCOE can be further reduced to~53.8~\euro/MWh. For a further discussion of these and related scenarios we refer the reader to~\cite{eriksen2017}.

\subsection{Flow tracing}
\label{sec:flow_tracing}
The technique of flow tracing allows to follow the power flows from the exporting source nodes through the network to the importing sink nodes. It was been introduced in parallel by Bialek et al. in~\cite{Bialek1996} and by Kirschen et al. in~\cite{Kirschen1997}. This method has for instance been proposed in~\cite{itc} as a flow allocation scheme in the context of the European inter transmission system operator compensation mechanism~(see the position paper~\cite{Acer2013} from the Agency of Energy Regulators for a discussion of this mechanism). It also has been used as an analytical tool for transmission capacity allocation in a simplified model of a highly renewable European electricity system in~\cite{Tranberg2015}. Here we will follow the extended formulation as presented in~\cite{Schaefer2017} and~\cite{Hoersch2017}, using a decomposition of power flows across the network into strands associated with the injecting export node and the type of generation. For this purpose we define the in-partition $q^{\text{in}}_{(n,\mu)}$, which describes the share of the total power $P_n^+ := \max(P_{n},0)$ injected at node $n$ associated with generation type $\mu\in\{W,S,B\}$ (wind, solar, backup power generation). This in-partition represents the input to the flow tracing algorithm, which then yields the following flow-partition and out-partition:
\begin{itemize}
\item The flow partition $\left\{ q_{l,(n,\mu)} \right\}$ dissects the flow $F_l$ on a link $l$ into the components $F_{l,(n,\mu)} = q_{l,(n,\mu)} F_l$ that are attributed to the exporting node $n$ and generation type $\mu$.
\item The out-partition $\left\{ q^\text{out}_{n,(m,\mu)} \right\}$ describes the composition of power $P_n^- := -\min(P_n,0)$ imported by node $n$. The respective share of this import $P_{n}^{-}$ attributed to the exporting node $m$ and generation type $\mu$ is given by $q^\text{out}_{n,(m,\mu)} P^-_n$.
\end{itemize}
The algorithm which determines this output from the in-partition is based on $(m,\mu)$-flow conservation and the principle of proportional sharing~(see the pioneering articles~\cite{Bialek1996} and~\cite{ Kirschen1997} for a discussion of this principle):
\begin{multline}\label{eq:flowtracing}
\delta_{n,m}q^{\text{in}}_{(n,\mu)}P^{+}_{n} +  \sum_{k}q_{k\rightarrow n,(m,\mu)}F_{k\rightarrow n}  \\
 =q^{\text{out}}_{n,(m,\mu)}P^{-}_{n} + \sum_{k}q_{n\rightarrow k,(m,\mu)}F_{n\rightarrow k}~,
\end{multline}
with
\begin{equation}
q_{n,(m,\mu)}^{\text{out}} = q_{n\to k,(m,\mu)}
\end{equation}
for the link $l=n\to k$ directed from node $n$ to node $k$. The links are oriented along the total power flow $F_{l}$. We solve~\eqref{eq:flowtracing} iteratively by starting at an exporting node without any inflow and following the power flow downstream. This is possible since the power flows on the network can be represented as a directed acyclic graph without loops, as discussed in~\cite{Schaefer2017}.

The flow tracing algorithm has to be applied to every instantaneous injection pattern $P_{n}(t)$ and corresponding power flow pattern $F_{l}(t)$. The resulting out-partitions $q_{n,(m,\mu)}^{\text{out}}$ can be integrated into the following average export transfer function $\mathcal{E}_{m\rightarrow n}^{\mu}$:
\begin{equation}\label{eq:export-transfer}
\mathcal{E}_{m\rightarrow n}^{\mu} =
\langle q_{n,(m,\mu)}^{\text{out}} P_{n}^{-}\rangle
~.
\end{equation}
This measure describes the average amount of power injected at node $m$ and resulting from generation type $\mu$, which is exported through the network to the importing node $n$.
\paragraph{Prioritization of intrinsic renewable generation and decomposition of nodal exports}\label{sec:prioritizing}
We use a simplified model of the European electricity network, in which each country is represented by a single node. By disregarding the internal transmission networks, we aggregate over power generation and consumption on smaller scales, and determine the corresponding coarse-grained  curtailment, backup power generation, imports, and exports on country scale. Nevertheless, since the flow tracing methodology considers the composition of the coarse-grained power flows and imports/exports, we have to introduce some additional rules on the nodal country level.

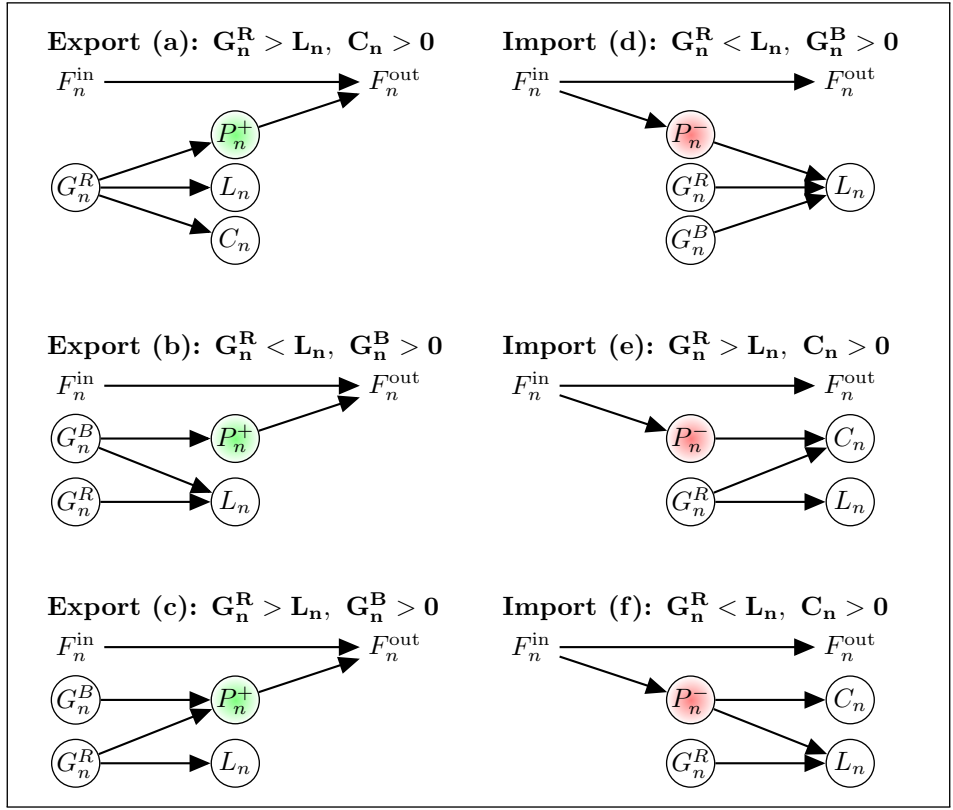
\begin{figure}[h!]
\centering
\vspace{-1em}
\setlength{\fboxsep}{2pt}%
\setlength{\fboxrule}{0.5pt}%
\fbox{
\begin{tabular*}{\textwidth}{c @{\extracolsep{\fill}}}
\vspace*{-0.5em}\\
\vspace{2.5em}
\begin{subfigure}{.44\textwidth}
  \normalsize{\textbf{Export (a): $\mathbf{G_{n}^{R} > L_{n}},\ \mathbf{C_n>0}$}}\\
  \begin{centering}
    \begin{tikzpicture}[scale=0.7]
      \SetGraphUnit{4}
      \node[rounded corners,inner sep=3pt,draw=white,fill=white] (p1) at (0,0) {$F_{n}^{\text{in}}$};
      \tikzset{VertexStyle/.append style={inner color=green!50}}
      \Vertex[L=$P^+_n$,x=3,y=-1]{p2};
      \tikzset{VertexStyle/.append style={inner color=white}}
      \node[rounded corners,inner sep=3pt,draw=white,fill=white] (p3) at (6,0) {$F_{n}^{\text{out}}$};
     	\Vertex[L=$G^R_n$,x=0,y=-2]{p4};
      \Vertex[L=$L_n$,x=3,y=-2]{p5};
      \Vertex[L=$C_n$,x=3,y=-3]{p6};
      \Edge(p1)(p3)
      \Edge(p2)(p3)
      \Edge(p4)(p2)
      \Edge(p4)(p5)
      \Edge(p4)(p6);
    \end{tikzpicture}
  \end{centering}
\end{subfigure}
\hspace{1.2em}
\begin{subfigure}{.44\textwidth}
  \normalsize{\textbf{Import (d): $\mathbf{G_{n}^{R} < L_{n}},\ \mathbf{G_n^B>0}$}}\\
  \begin{centering}
    \begin{tikzpicture}[scale=0.7]
      \SetGraphUnit{4}
      \node[rounded corners,inner sep=3pt,draw=white,fill=white] (p1) at (0,0) {$F_{n}^{\text{in}}$};
      \tikzset{VertexStyle/.append style={inner color=red!50}}
      \Vertex[L=$P^-_n$,x=3,y=-1]{p2};
      \tikzset{VertexStyle/.append style={inner color=white}}
      \node[rounded corners,inner sep=3pt,draw=white,fill=white] (p3) at (6,0) {$F_{n}^{\text{out}}$};
      \Vertex[L=$G^R_n$,x=3,y=-2]{p4};
      \Vertex[L=$L_n$,x=6,y=-2]{p5};
      \Vertex[L=$G^B_n$,x=3,y=-3]{p6};
      \Edge(p1)(p2)
      \Edge(p1)(p3)
      \Edge(p2)(p5)
      \Edge(p4)(p5)
      \Edge(p6)(p5);
    \end{tikzpicture}
  \end{centering}
\end{subfigure}\\

\vspace{2.5em}
\begin{subfigure}{.44\textwidth}
  \normalsize{\textbf{Export (b): $\mathbf{G^R_n < L_n,\ G^B_n>0}$}}\\
  \begin{centering}
    \begin{tikzpicture}[scale=0.7]
      \SetGraphUnit{4}
      \node[rounded corners,inner sep=3pt,draw=white,fill=white] (p1) at (0,0) {$F_{n}^{\text{in}}$};
      \tikzset{VertexStyle/.append style={inner color=green!50}}
      \Vertex[L=$P^+_n$,x=3,y=-1]{p2};
      \tikzset{VertexStyle/.append style={inner color=white}}
      \node[rounded corners,inner sep=3pt,draw=white,fill=white] (p3) at (6,0) {$F_{n}^{\text{out}}$};
      \Vertex[L=$G^B_n$,x=0,y=-1]{p4};
      \Vertex[L=$L_n$,x=3,y=-2.2]{p5};
      \Vertex[L=$G^R_n$,x=0,y=-2.2]{p6};
      \Edge(p1)(p3)
      \Edge(p2)(p3)
      \Edge(p4)(p2)
      \Edge(p4)(p5)
      \Edge(p6)(p5);
    \end{tikzpicture}
  \end{centering}
\end{subfigure}
\hspace{1.2em}
\begin{subfigure}{.44\textwidth}
  \normalsize{\textbf{Import (e): $\mathbf{G^R_n>L_n,\ C_n>0}$}}\\
  \begin{centering}
    \begin{tikzpicture}[scale=0.7]
      \SetGraphUnit{4}
      \node[rounded corners,inner sep=3pt,draw=white,fill=white] (p1) at (0,0) {$F_{n}^{\text{in}}$};
      \tikzset{VertexStyle/.append style={inner color=red!50}}
      \Vertex[L=$P^-_n$,x=3,y=-1]{p2};
      \tikzset{VertexStyle/.append style={inner color=white}}
      \node[rounded corners,inner sep=3pt,draw=white,fill=white] (p3) at (6,0) {$F_{n}^{\text{out}}$};
      \Vertex[L=$G^R_n$,x=3,y=-2.2]{p4};
      \Vertex[L=$C_n$,x=6,y=-1]{p5};
      \Vertex[L=$L_n$,x=6,y=-2.2]{p6};
      \Edge(p1)(p2)
      \Edge(p1)(p3)
      \Edge(p2)(p5)
      \Edge(p4)(p5)
      \Edge(p4)(p6);
    \end{tikzpicture}
  \end{centering}
\end{subfigure}\\

\vspace{0.5em}
\begin{subfigure}{.44\textwidth}
  \normalsize{\textbf{Export (c): $\mathbf{G^R_n>L_{n},\ G^B_n>0}$}}\\
  \begin{centering}
    \begin{tikzpicture}[scale=0.7]
      \SetGraphUnit{4}
      \node[rounded corners,inner sep=3pt,draw=white,fill=white] (p1) at (0,0) {$F_{n}^{\text{in}}$};
      \tikzset{VertexStyle/.append style={inner color=green!50}}
      \Vertex[L=$P^+_n$,x=3,y=-1]{p2};
      \tikzset{VertexStyle/.append style={inner color=white}}
      \node[rounded corners,inner sep=3pt,draw=white,fill=white] (p3) at (6,0) {$F_{n}^{\text{out}}$};
      \Vertex[L=$G^B_n$,x=0,y=-1]{p4};
      \Vertex[L=$L_n$,x=3,y=-2.2]{p5};
      \Vertex[L=$G^R_n$,x=0,y=-2.2]{p6};
      \Edge(p1)(p3)
      \Edge(p2)(p3)
      \Edge(p4)(p2)
      \Edge(p6)(p2)
      \Edge(p6)(p5);
    \end{tikzpicture}
  \end{centering}
\end{subfigure}
\hspace{1.2em}
\begin{subfigure}{.44\textwidth}
  \normalsize{\textbf{Import (f): $\mathbf{ G^R_n<L_n,\ C_n>0}$}}\\
  \begin{centering}
    \begin{tikzpicture}[scale=0.7]
    	\SetGraphUnit{4}
      \node[rounded corners,inner sep=3pt,draw=white,fill=white] (p1) at (0,0) {$F_{n}^{\text{in}}$};
      \tikzset{VertexStyle/.append style={inner color=red!50}}
      \Vertex[L=$P^-_n$,x=3,y=-1]{p2};
      \tikzset{VertexStyle/.append style={inner color=white}}
      \node[rounded corners,inner sep=3pt,draw=white,fill=white] (p3) at (6,0) {$F_{n}^{\text{out}}$};
     	\Vertex[L=$G^R_n$,x=3,y=-2.2]{p4};
      \Vertex[L=$C_n$,x=6,y=-1]{p5};
      \Vertex[L=$L_n$,x=6,y=-2.2]{p6};
      \Edge(p1)(p2)
      \Edge(p1)(p3)
      \Edge(p2)(p5)
      \Edge(p2)(p6)
      \Edge(p4)(p6);
    \end{tikzpicture}
  \end{centering}
\end{subfigure}
\end{tabular*}}
\caption{Nodal electricity partition cases. For every hour and every node one of these cases applies. Cases a-c all apply to nodal exports ($P^+_n=P_{n}>0$), while cases d-f all correspond to nodal imports ($P^-_n=-P_{n}>0$).}
\label{fig:partition-cases}
\end{figure}

Recall that the total net injection $P_{n}(t)$ of node $n$ is determined by the nodal balancing equation~\eqref{eq:balancing}:
\begin{equation}
P_{n}(t) = G_{n}^{R}(t) - L_{n}(t) + G_{n}^{B}(t) - C_{n}(t)~.
\end{equation}
Here we have written the balancing $B_{n}(t)$ in terms of backup power generation $G_{n}^{B}(t)$ and curtailment $C_{n}(t)$. Both the renewable generation $G_{n}^{R}(t)$ and nodal load $L_{n}(t)$ are given by the respective time series, whereas the balancing is determined from the heuristic scheme introduced in~\eqref{eq:synchronized_balancing}. Although all terms in this equation are uniquely defined, the composition of exports $P_{n}^{+}$ or load $L_{n}(t)$ can be interpreted in different ways. Consider in particular an exporting node with $P_{n}>0$, which besides renewable generation $G_{n}^{R}(t)$ provides backup power generation $G_{n}^{B}(t)$. In this case, which part of the renewable and backup power generation is assigned to the load $L_{n}(t)$ and which part to the net injection $P^{+}_{n}(t)$? In order to clarify this and related issues, in~\fref{fig:partition-cases} we distinguish between six different nodal electricity partition cases.

For the stated example, we apply a prioritization of intrinsic renewable generation, that is we assign as much intra-nodal renewable power generation as possible to the respective load, whereas a maximum of possible backup power generation is exported. Inside the boundaries of the simplified model considered here, this approach assumes that the individual countries use their generation capacity to their own benefit by exporting more expensive backup power generation from gas turbines, while using cheaper renewable generation for their own consumption. Note that the applied dispatch scheme distributes both backup power generation and curtailment to all countries relative to their average load.

The different nodal electricity partition cases illustrated in~\fref{fig:partition-cases} determine uniquely the in-partition $q_{(n,\mu)}$ for the flow tracing algorithm, and are consistent with a prioritization of intrinsic renewable generation:\\
\medskip
\begin{subequations}
\allowdisplaybreaks
\label{eq:inpartitions}
\textbf{Case (a):}
\begin{align}
    q_{(n,\mu)}^\text{in}(t) = \begin{dcases}
    \frac{G^W_n(t)}{G^R_n(t)}& \text{for }\mu=W\\
    \frac{G^S_n(t)}{G^R_n(t)}& \text{for }\mu=S\\
    0 & \text{for }\mu=B \end{dcases}~.
\end{align}
\textbf{Case (b):}
\begin{align}
    q_{(n,\mu)}^\text{in}(t) = \begin{dcases} 0 & \text{for }\mu=W\\ 0 & \text{for }\mu=S\\ 1 & \text{for }\mu=B \end{dcases}~.
\end{align}
\textbf{Case (c):}
\begin{align}
    q_{(n,\mu)}^\text{in}(t) = \begin{dcases}
    \frac{G^W_n(t)}{G^R_n(t)} \frac{G^R_n(t) - L_n(t)}{G^B_n(t) + G^R_n(t) - L_n(t)}& \text{for }\mu=W\\
    \frac{G^S_n(t)}{G^R_n(t)} \frac{G^R_n(t) - L_n(t)}{G^B_n(t) + G^R_n(t) - L_n(t)}
    & \text{for }\mu=S\\ \frac{G^B_n(t)}{G^B_n(t)+G^R_n(t)-L_n(t)} & \text{for }\mu=B \end{dcases}~.
\end{align}
\textbf{Cases (d), (e), (f):}
\begin{align}
    q_{(n,\mu)}^\text{in}(t) = \begin{dcases} 0 & \text{for }\mu=W\\ 0 & \text{for }\mu=S\\ 0 & \text{for }\mu=B \end{dcases}~.
\end{align}
\end{subequations}
Here, (\ref{eq:inpartitions}a) refers to case (a) in \fref{fig:partition-cases} and so forth. Note that only cases (a), (b), and (c) lead to a non-zero in-partition.

\paragraph{Transmission network usage measure}\label{sec:usage-measure}
The system participants make use of the transmission infrastructure through the power flows resulting from their imports and exports. The respective infrastructure costs thus have to be allocated to these users based on a suitable network usage measure. Although such a measure could be derived from nodal properties only, like average imports and exports (postage stamp method), this approach would ignore the role of the individual nodes in the fluctuating spatio-temporal flow pattern. In the following we apply a flow-based transmission capacity allocation measure, which has been introduced in~\cite{Tranberg2015}.  Using the flow tracing method, this measure attributes the transmission capacity of a link $\mathcal{K}_{l}^T$, defined in \eqref{eq:transmission_capacity}, to each country $n$ based on the statistics of the corresponding partial flows $F_{l,n}(t)=q_{l,n}(t)F_{l}(t)$:
\begin{equation}
    \label{eq:transcapusage}
    \mathcal{K}_{l,n}^T = \int_0^{\mathcal{K}_l^T}d\mathcal{K}\frac{1}{1-P_l(\mathcal{K})} \int_\mathcal{K}^{\mathcal{K}_l^T}dF_lp_l(F_l)\langle q_{l,n}|F_{l}\rangle~.
\end{equation}
Here $p_l$ is the probability distribution of the flow on the link $l$, $P_l$ is the cumulative probability function, and $\langle q_{l,n}|F_{l}\rangle$ is the average fraction of the flow assigned to the source node $n$, given that the total flow is equal to $F_{l}$. Note that we here do not discriminate different generation technologies and thus omit the index~$\mu$. In~\cite{Tranberg2015} this measure is derived by virtually decomposing the total capacity $\mathcal{K}^T_{l,n}$ into infinitesimal increments $d\mathcal{K}$ for the whole time series. Each of these increments is then weighted according to the respective flow-based usage by the node $n$. The structure of the expression~\eqref{eq:transcapusage} assures that the first increments of the total capacity are assigned to all users with partial flows on the respective link, whereas the last increments close to the total capacity are only assigned to those system participants which make use of the link in high-flow situations. For a further discussion of this approach we refer the reader to~\cite{Tranberg2015} and~\cite{Hoersch2017}. The detailed treatment of an analytical test case in~\cite{Schaefer2017} also illustrates the concept underlying the measure in~\eqref{eq:transcapusage}.
\section{Flow-based nodal cost allocation}
\label{sec:nodal_cost_allocation}
\subsection{Nodal cost allocation}
The system LCOE \eqref{eq:system-lcoe} is an aggregated measure of the cost of electricity across all countries. Allocating costs for individual nodes requires a nodal LCOE measure. Here we introduce such a measure with the requirement that the weighted sum of nodal costs reproduces the aggregated system costs
\begin{equation}\label{eq:lcoe-decomp}
    \text{LCOE}_\text{EU} = \sum_n \frac{\mean{L_n}}{\mean{L_\text{EU}}} \text{LCOE}_n \,.
\end{equation}
The costs related to the individual system components in \eqref{eq:present_value} and \eqref{eq:trans-value}, constituting the LCOE \eqref{eq:lcoe}, have to be allocated to the respective countries. This corresponds to the introduction of a node index on the set of present values:
\begin{equation}
\{V_W, V_S, V_B, V_T\} \Rightarrow \{V_{W,n}, V_{S,n}, V_{B,n}, V_{T,n}\}.
\end{equation}
In the following we differentiate between generation capacity costs and transmission infrastructure costs. A straightforward method for the attribution of generation capacity costs would be according to their respective geographical location, that is for instance the cost of wind power generation capacity is assigned to the country in which the respective turbines are installed. Such an allocation ignores the fact that in an interconnected system with transmission a significant amount of generated power might be exported to and thus consumed in another country. In order to establish nodal LCOEs which take into account these import/export patterns, we thus have to connect the location of generation with the location of consumption. This connection is realized by the flow tracing method and expressed by the export transfer function $\mathcal{E}_{m\rightarrow n}^{\mu}$. Based on this function, we define the measure $\mathcal{K}_{m\rightarrow n}^{\mu}$, that describes the generation capacity of type $\mu\in\{W,S,B\}$ located at node $m$ which is used by node $n$:
\begin{equation}
\label{eq:capacity_assignment}
\mathcal{K}_{m\rightarrow n}^{\mu} =
\begin{cases}
\left[\frac{\mathcal{E}_{m\rightarrow n}^{\mu}}{
\langle G_{m}^{\mu}\rangle - \langle C_{m}^{\mu}\rangle
}\right]
\mathcal{K}_{m}^{\mu} & \mbox{if } m \neq n\\
\mathcal{K}_{m}^{\mu} - \sum_{s\neq m} \mathcal{K}_{m\rightarrow s}^{\mu} & \mbox{if } m=n~.
\end{cases}
\end{equation}
Here $\langle G_{m}^{\mu}\rangle $ and $\langle C_{m}^{\mu}\rangle$ denote the average generation and curtailment of generation type $\mu$ in node $m$, respectively. In the case $m\neq n$ this measure attributes a share  $\mathcal{K}_{m}^{\mu}$ of the generation capacity of type $\mu$ in node $m$ to an importing node $n$. The attribution is proportional to the respective export transfer function $\mathcal{E}_{m\rightarrow n}^{\mu}$ and the amount of power of type $\mu$ generated at node $m$ that is not curtailed. The capacity that gets attributed to node $m$ itself is the remaining capacity after the attribution to all importing nodes $n$. With this measure we can calculate the total share of system generation capacity of type $\mu$ which is attributed to node $n$:
\begin{equation}\label{eq:nodal-capacity}
\widetilde{\mathcal{K}}_{n}^{\mu} = \sum_{m}\mathcal{K}_{m\rightarrow n}^{\mu}~.
\end{equation}
In a similar way the share of the total system backup energy generation is attributed to a node $n$ by:
\begin{align}\label{eq:nodal-energy}
\widetilde{E}_{n}^{B} &=
\langle G_{n}^{B}\rangle - \sum_{m\neq n} \mathcal{E}_{n\rightarrow m}^{B}
+\sum_{m\neq n}\mathcal{E}_{m\to n}^{B}\\
&= \sum_{m}\frac{\mathcal{K}^{B}_{m\rightarrow n}}{\mathcal{K}_{m}^{B}} \langle G_{m}^{B} \rangle~.
\end{align}
Here we have used the definition in~\eqref{eq:capacity_assignment} and $\langle C_{m}^{B}\rangle=0$. Combining the nodal generation capacities \eqref{eq:nodal-capacity} and the nodal system backup energy usage \eqref{eq:nodal-energy}, the nodal LCOE is calculated as
\begin{equation}
\label{eq:nodal_lcoe2}
 \text{LCOE}_{n} = \sum_{\mu}\frac{\widetilde{V}^\mu_n}{
\sum_{t=1}^{T^\mu_\text{life}}\frac{L_{n,t}}{(1+r)^{t}}
} + \frac{V^T_n}{
\sum_{t=1}^{T_{\text{life}}^T}\frac{L_{n,t}}{(1+r)^{t}}
}~,
\end{equation}
where the nodal present values of investment $\widetilde{V}^{\mu}_{n}$ are now based on $\widetilde{\mathcal{K}}_{n}^{\mu}$ and $\widetilde{E}_{n}^{B}$ where applicable. For the allocation of transmission capacity costs $V_{n}^{T}$ in the nodal LCOEs we consider different schemes. A simple realization based on nodal properties only is an assignment according to the average load of a country:
\begin{equation}
\label{eq:transcapcost_meanload}
V_{n}^{T}=\frac{\langle L_{n}\rangle}{\langle L_{\text{EU}}\rangle} V^{T}~.
\end{equation}
An alternative flow-based transmission infrastructure cost allocation makes use of the capacity usage measure $\mathcal{K}_{l,n}^{T}$ in~\eqref{eq:transcapusage}:
\begin{equation}
\label{eq:transcapcost_flowbased}
V^T_n = \sum_{l}\mathcal{K}_{l,n}^{T}d_{l}c_{l}~.
\end{equation}
Here we assume that the cost for the DC converter stations are allocated following the assignment of the corresponding HVDC lines. Note that $\mathcal{K}_{l,n}^{T}$ can be either based on exports or imports, depending on how the flow tracing algorithm is applied.
\subsection{Variation of nodal costs}
Our primary goal with the introduction of heterogeneity is to reduce the overall system costs. As a side effect, the introduced heterogeneity of the renewable generation capacities might lead to a change in the variation of the individual nodal costs. To measure this change, we use the weighted standard deviation (WSD) of the nodal LCOEs:
\begin{equation}\label{eq:wsd}
    \text{WSD} = \sqrt{\sum_{n=1}^N \frac{\langle L_n \rangle}{\langle L_\text{EU} \rangle} \left(\text{LCOE}_n - \text{LCOE}_\text{EU}\right) ^2}\;.
\end{equation}
\section{Results}
\label{sec:results}
\subsection{Patterns of imports and exports}
\begin{figure}[h!] \centering
\includegraphics[width = \columnwidth]{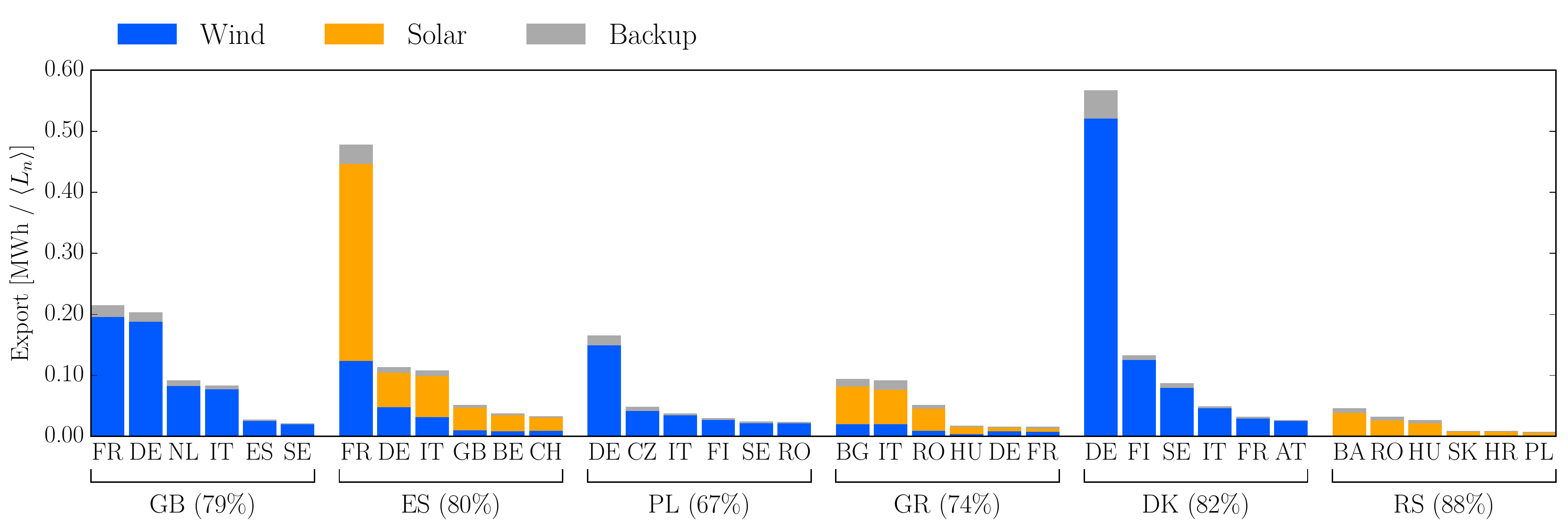}\\
\includegraphics[width = \columnwidth]{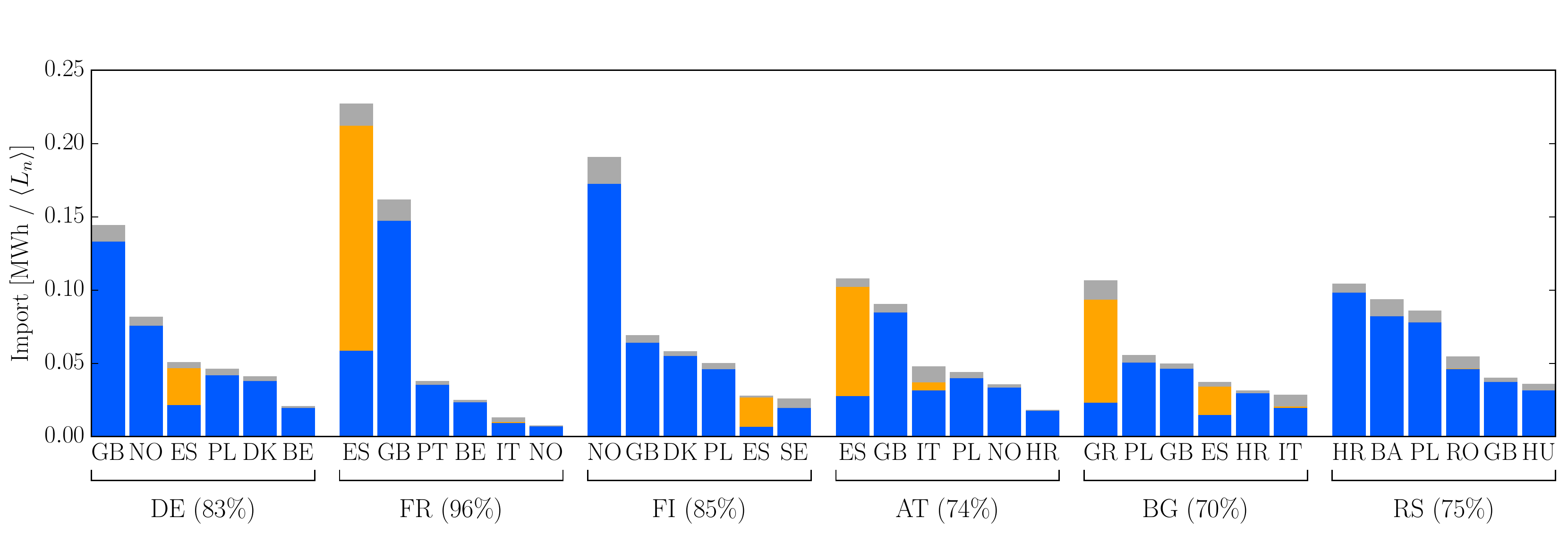}
\caption{Average exports (top) and imports (bottom) for selected countries in the GAS* K=2 layout. Results are calculated using the export \eqref{eq:export-transfer} transfer function. The percentage in parenthesis denotes the amount of the country's total average export (top) or import (bottom) accounted for by the sum of the six bars.}
\label{fig:import-export}
\end{figure}

We now look at the patterns of imports and exports for the GAS* K=2 layout. In the top panel of \fref{fig:import-export} we show the six largest average exports for six countries, which are all, except for Serbia (RS), net exporters with $\gamma_n>1$; see \fref{fig:gamma_layout} for reference. The average exports are calculated using the export transfer function \eqref{eq:export-transfer} and displayed in terms of the mean load of the country of interest. The percentage in parenthesis denotes the amount of the country's total average export accounted for by the sum of the six bars. Similarly, the bottom panel shows the six largest average imports for six countries, all net importers with $\gamma_n<1$.

The leftmost part of the top panel displays the top six countries that Great Britain is exporting to (normalized to the mean load of Great Britain). It shows that 79\% of the average exports of Great Britain are accounted for by only six countries. Similarly, the leftmost part of the bottom panel shows the top six countries contributing to the imports of Germany (normalized to the mean load of Germany). In both panels the six countries of interest are sorted by descending mean load.

The mix between wind and solar, which is shown in \fref{fig:gamma_layout}, is visible in the export components in the top panel, with GB and DK having only wind and ES having a solar dominated mix.

All net importers in the bottom panel are importing from GB, and almost all are importing from ES as well. This holds even for countries far away in the network, such as FI, BG, RS, see \fref{fig:map} for reference. GB and ES have considerable mean load and $\gamma_n$ close or equal to 2, which is why their exports are ubiquitous in the network. Note that the renewable mix of Spain is solar dominated while GB is purely wind.

Serbia is included in both panels of~\fref{fig:import-export}. It is a net importer with $\gamma_n=0.5$. We see that even net importers will be exporting during some hours. Since for Serbia $\alpha_{n}=0$, solar and backup power generation accounts for the entire export. The imports of Serbia are almost entirely wind, which is due to the dominance of wind power generation in the net exporting neighboring countries.

\subsection{Nodal cost of electricity}
\begin{figure}[h!]
 \centering
\includegraphics[width = \columnwidth]{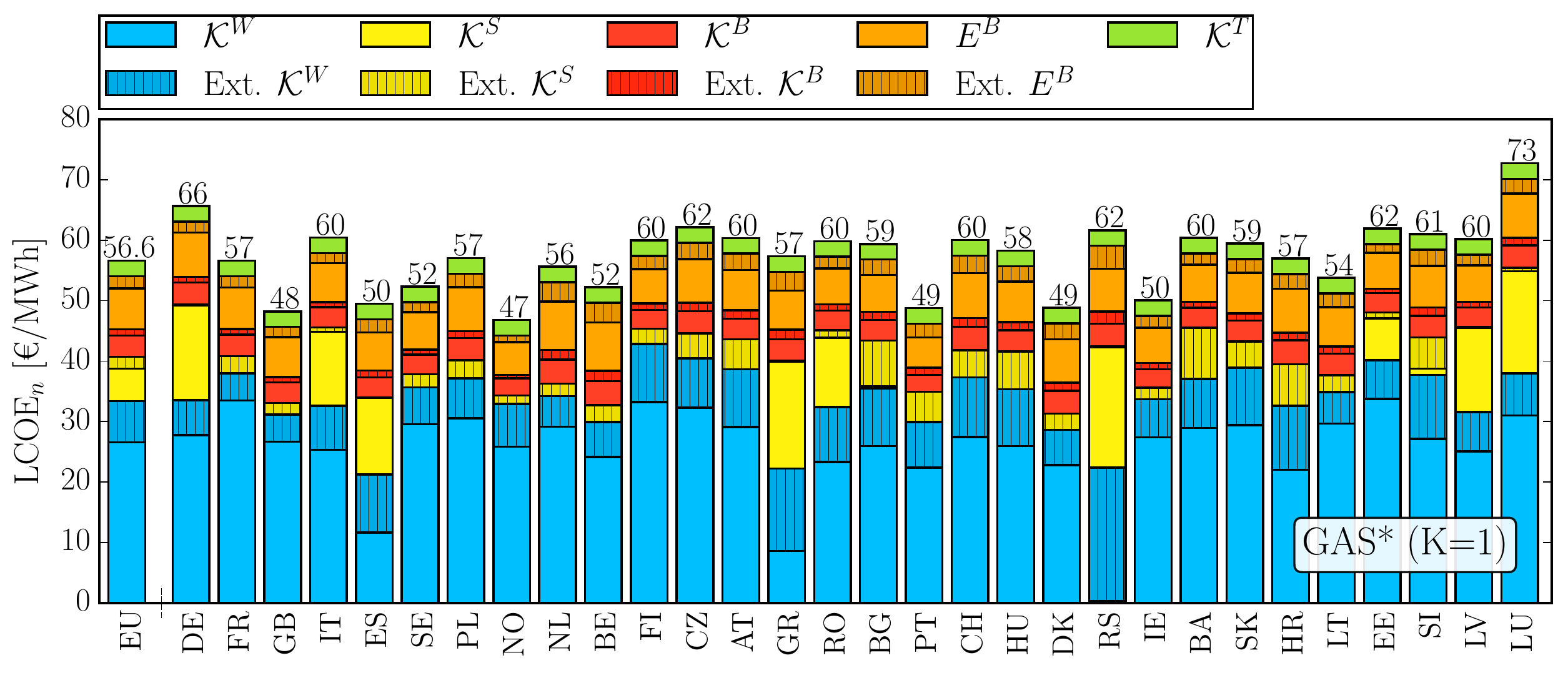}\\
\includegraphics[width = \columnwidth]{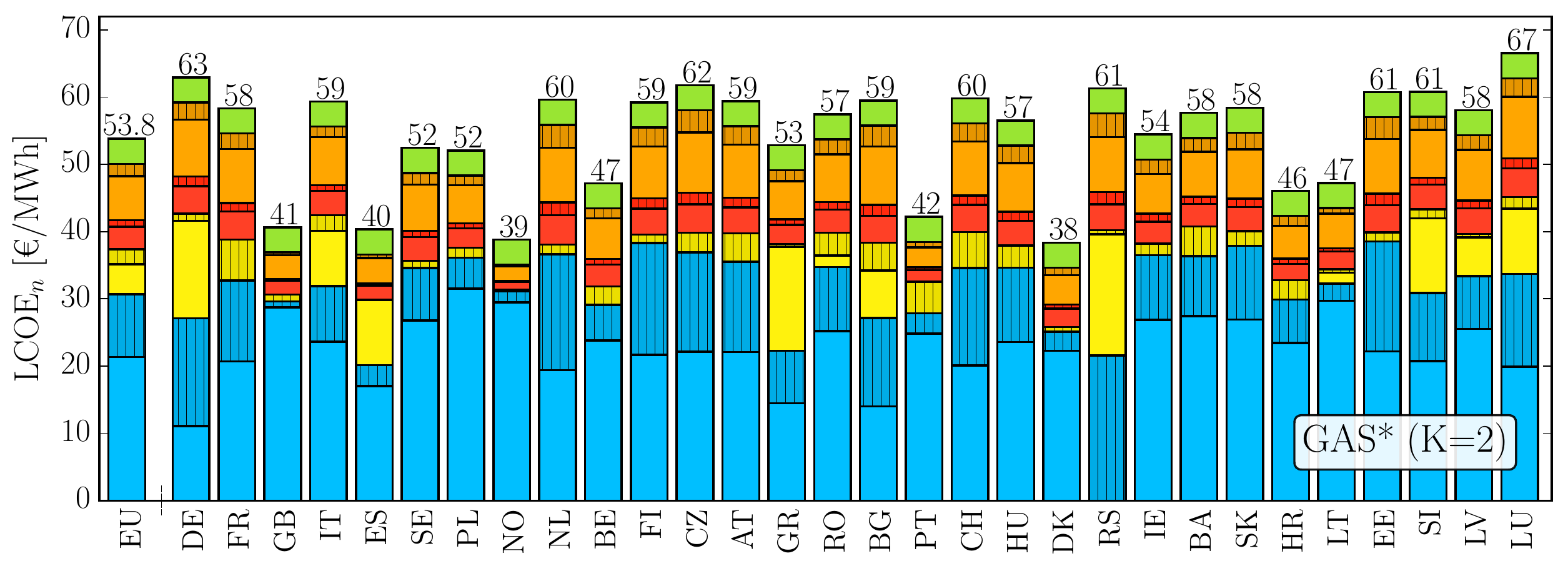}
\caption{Nodal component-wise LCOE for the GAS* K=1 (top) and the GAS* K=2 (bottom) layout. The components are split into their local and external parts. In both cases the transmission cost component of the LCOE is distributed according to the average load of the individual countries as defined in~\eqref{eq:transcapcost_meanload}.}
\label{fig:nodal_lcoe}
\end{figure}

We apply the method of flow-based nodal LCOEs as defined by~\eqref{eq:nodal_lcoe2} to the two GAS* optimized scenarios with transmission and heterogeneity limits K=1 and K=2 (see ~\sref{sec:model}). For the heterogeneity constraint of K=1 the results are shown in the top part of \fref{fig:nodal_lcoe}, whereas the bottom part shows the results for the more heterogeneous layout for K=2. In these figures, the transmission capacity costs are attributed equally to the countries based on their respective mean load according to~\eqref{eq:transcapcost_meanload}.
We observe that the external capacity and backup energy costs allocated using the flow tracing method represent a significant part of the nodal LCOE. This becomes particularly apparent for Serbia, which is assigned almost none local wind power capacity (see \fref{fig:gamma_layout}), but uses a substantial amount of external wind power generation to cover its electricity demand; see also \fref{fig:import-export}. Similarly, for both layouts with transmission France has no local solar generation capacity, but to some extent imports such solar power generation through the transmission grid. The usage of backup power capacity and backup energy is relatively homogeneously distributed for most nodes, which has its origin in the synchronized balancing scheme (recall the definition in~\eqref{eq:synchronized_balancing}).

\begin{figure}[h!]
\centering
\includegraphics[width = \columnwidth]{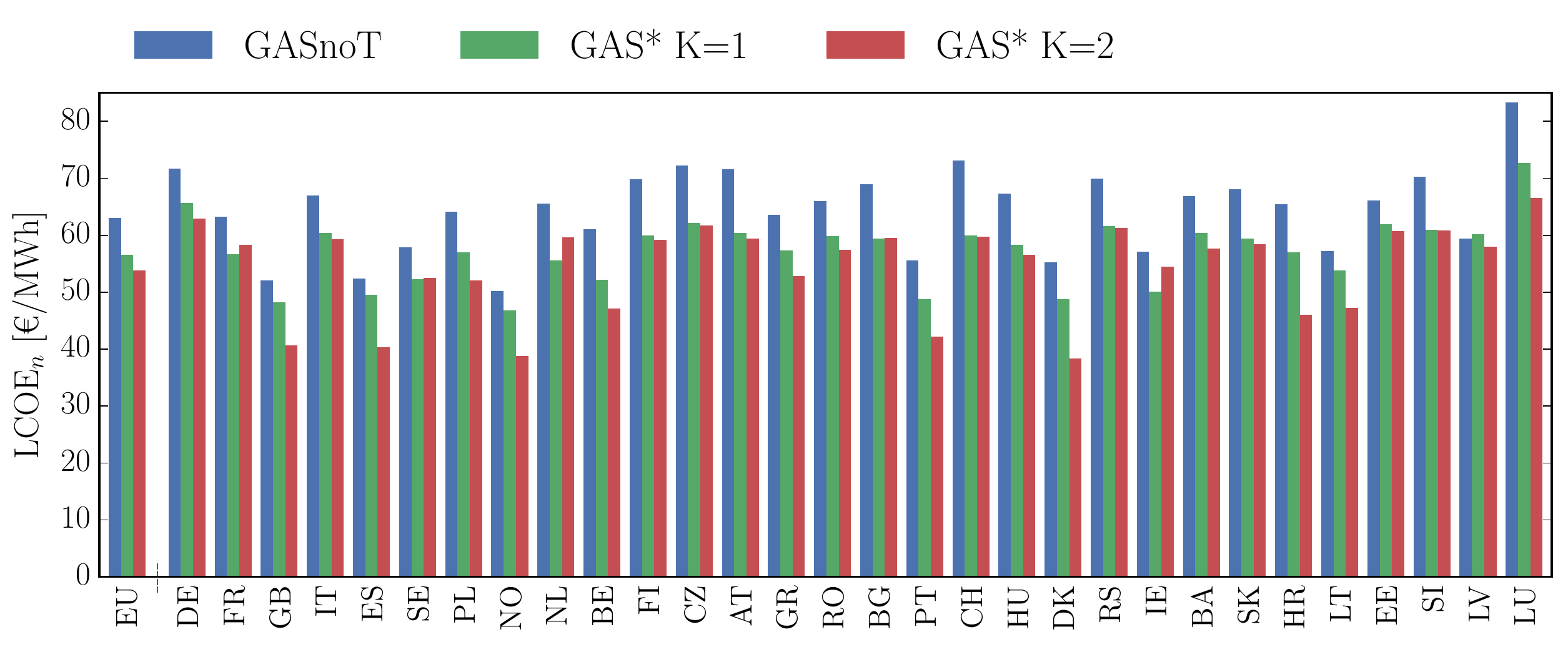}
\caption{Comparison of nodal LCOEs for GASnoT (blue), GAS* K=1 (green), and GAS* K=2 (red). Allowing transmission reduces the system LCOE and all nodal LCOEs. Increasing the heterogeneity from K=1 to K=2 leads to a decrease of the system LCOE and all nodal LCOEs except small increases for FR, SE, NL, BG and IE.}
\label{fig:nodal_lcoe_comparison}
\end{figure}

In~\cite{eriksen2017} it was shown that power transmission between the European countries and a higher degree of heterogeneity lowers the total system LCOE from 63\euro/MWh to~53.8\euro/MWh. This is shown in the leftmost bar of~\fref{fig:nodal_lcoe_comparison}. How does this cost reduction translate to the nodal LCOEs associated with the individual countries? Remarkably, the transition from the GAS~K=1 layout without transmission to the GAS*~K=1 layout with transmission leads to reduced nodal LCOE for every single country; see \fref{fig:nodal_lcoe_comparison}. This shows that under a flow-based cost allocation mechanism, cooperation in a European electricity system through power transmissions does not only lead to a global, but also to a system-wide local cost reduction. Allowing a more heterogeneous GAS* K=2 layout decreases the system LCOE further from 56.6\euro/MWh to 53.8\euro/MWh. Apart from a few exceptions (France, Sweden, Netherlands, Bulgaria, Ireland) this global cost reduction is again related with reduced or unchanged LCOEs on the nodal level. Compared with the system layout without transmission, we observe the largest drop in nodal LCOEs for Denmark (DK) and Luxembourg (LU), with a nodal cost reduction of~17\euro/MWh and~16\euro/MWh, respectively, compared to an overall cost reduction of~9.2\euro/MWh. In the case of Luxembourg this cost reduction is due to its low capacity factor for local wind energy~\cite{eriksen2017}, whereas for Denmark the cost reduction can be explained in an efficient usage of abroad generation capacity combined with a high local capacity factor for wind energy.

In both panels of \fref{fig:nodal_lcoe}, we see fluctuations in nodal LCOEs around the EU average shown on the left. In~\tref{tab:wsd} we display the WSD as defined in~\eqref{eq:wsd} as a measure of the heterogeneity of nodal LCOEs for the scenario without transmission and for the scenarios with transmission (heterogeneity parameter K=1 and K=2). We observe that K=1 the incorporation of power transmission reduces the WSD from~7.27~\euro/MWh to~5.38~\euro/MWh. This shows that the spread in nodal LCOEs is at first reduced by cooperation between the European countries, which can be explained by the partial smoothing out of the heterogeneity in the renewable generation resources and the accompanying generation costs. For increasing heterogeneity K=2 we then observe an increasing WSD from 5.38~\euro/MWh to 8.54~\euro/MWh. This shows that even under the flow based nodal cost allocation scheme the system-wide cost reduction due to higher heterogeneity is unequally distributed among the different countries. In particular, net exporting countries with $\gamma_{n}>1$ are the main beneficiaries of the transition from the K=1 to the K=2 scenario (see~\fref{fig:nodal_lcoe_comparison}). These countries are able to consume higher shares of renewable generation, therefore saving costs in backup power generation, while exporting the involved surplus and the related capacity costs. However, this benefit from heterogeneity goes along with a disproportional usage of the transmission grid, which is not represented in the transmission cost allocation based on average loads applied so far. In the next subsection we explore how a respective flow based allocation affects the spread in the nodal LCOEs.

\subsection{Transmission cost allocations}
\begin{figure}[h!] \centering
\includegraphics[width = \columnwidth]{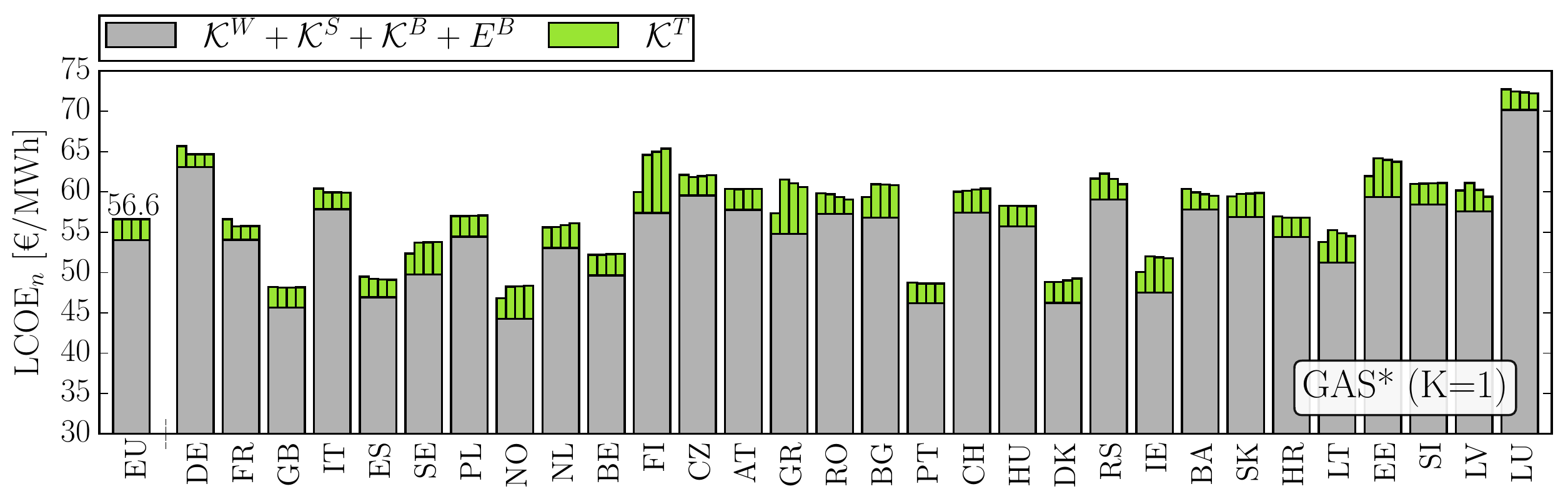}\\
\includegraphics[width = \columnwidth]{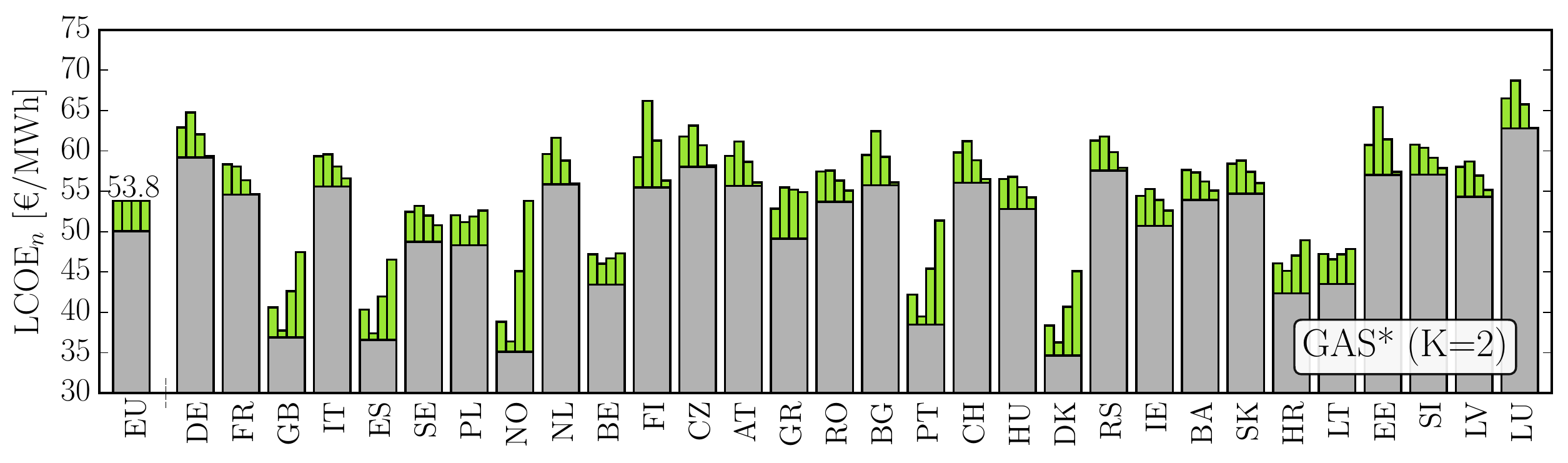}
\caption{Comparison of four transmission cost allocations. Leftmost of the green bars: assignment proportional to average loads. Then from left to right: importing countries alone pay for the transmission line costs, equal sharing of expenses between both importer and exporter, and transmission cost allocated to exporting countries. Note that all transmission cost allocations will always result in the same weighted average value of the total European LCOE.}
\label{fig:nodal_lcoe_trans}
\end{figure}

Besides a transmission cost allocation according to the average load of a country, in the following we consider three variants of the flow based cost assignment in~\eqref{eq:transcapcost_flowbased}: 1) the importer alone pays for transmission, 2) the transmission expenses are shared equally between both importer and exporter, and 3) the exporter alone pays for the transmission. The respective underlying network usage measure~$\mathcal{K}_{l,n}$ in~\eqref{eq:transcapusage} in general is used to allocate transmission capacity to exporters by tracing the exported power flows. This is the export picture, which is described in~\sref{sec:flow_tracing} and is used for case 3). To allocate transmission capacity to importers we switch to the import picture, that is $P_n(t)\to -P_n(t)$ and $F_{l}(t)\to -F_{l}(t)$, and apply \eqref{eq:transcapusage} to trace the imported power flows for case 1)~\cite{Tranberg2015}. Case 2) is an average of the import and export picture. These three cases of flow based transmission cost allocation are compared with the one based on average loads in \fref{fig:nodal_lcoe_trans}. The nodal LCOEs values are, except for the transmission component, equal to the results shown in~\fref{fig:nodal_lcoe}. For this reason we highlight the transmission cost component of the nodal LCOEs in~\fref{fig:nodal_lcoe_trans}, leaving the other components grey. The four green bars shown for each country represent from left to right the four cases described above.

The top panel of~\fref{fig:nodal_lcoe_trans} shows results for the homogeneous GAS* K=1 layout with $\gamma_n=1$, while the bottom panel shows results for the more heterogeneous GAS* K=2 layout with $0.5\leq \gamma_n\leq 2$. For the K=1 layout for all node there are only minor differences between the three cases of transmission cost allocation. This is due to the fact that, on average, each country serves its own load by its own renewable generation, and only the fluctuating surpluses are subject to power transmission. In the more heterogeneous case of K=2, however, we see significant differences between the three transmission cost allocations. In this scenario the occurrence of net importers and net exporters lead to a more disproportionate usage of the transmission grid. As discussed in the previous section, the transition from the K=1 to the K=2 scenario mainly benefits the net exporting countries. Accordingly we observe in~\tref{tab:wsd} an increasing WSD if we allocate transmission costs using the flow based methodology based on imports. The equal partition between exports and imports then reduces the WSD compared to the case of a transmission cost assignment based on average loads due to its incorporation of some of the exporters significant transmission usage. Although such an allocation scheme could be considered as most fair by representing the individual countries' role in the overall flow pattern, it does not balance the benefit of higher heterogeneity in terms of nodal LCOEs for net exporters. Accordingly, the WSD in~\tref{tab:wsd} as expected adopts its lowest value in the case of a purely export based transmission cost allocation. Such an attribution mainly affects the six countries with $\gamma_n$ close or equal to 2, which incur the strongest benefit from the increasing heterogeneity in the system. The respective countries are Great Britain, Spain, Norway, Portugal, Denmark and Croatia (see~\fref{fig:gamma_layout} for reference), which as shown in~\fref{fig:nodal_lcoe_trans} display disproportionally large allocations of transmission infrastructure costs.

\begin{table}[htp!]
\centering
\caption{The weighted standard deviation (WSD) of the fully usage-based nodal LCOE for different shares of the transmission cost between the importing and exporting countries. Results are shown for different values of the constraint parameter $K$ and different flow allocation schemes. The LCOE$_\text{EU}$ values are shown for reference. All results are obtained for the GAS* layouts and all values are expressed in units of \euro/MWh.}
\label{tab:wsd}
\begin{tabular}{l|c|r|rcc}
\toprule
\textbf{Heterogeneity} & \textbf{LCOE$_\text{EU}$} & \textbf{WSD} & \multicolumn{3}{l}{\textbf{WSD for import/export:}}\\
\textbf{constraint}  &   &           & \textbf{100/0} & \textbf{50/50} & \textbf{0/100} \\
\midrule
$K\!=1$ {\small(no trans.)} & 63.0 & 7.27     & & & \\
$K\!=1$ & 56.6 & 5.38     & 5.28                & 5.22   & 5.17   \\
$K\!=2$ & 53.8 & 8.54     & 10.22 & 6.72   & 3.83   \\
\bottomrule
\end{tabular}
\end{table}
\section{Conclusion and outlook}
\label{sec:conclusion}
In this contribution we study flow-based nodal levelized costs of electricity (LCOEs) in the framework of a simplified model of a highly renewable European electricity network with countries as aggregated network nodes. As shown in~\cite{eriksen2017}, an optimal heterogeneous placement of renewable generation capacity reduces the system-wide LCOE in this model. In order to be able to investigate LCOEs on a nodal level, we use a flow tracing technique to connect the location of power generation with the corresponding location of consumption. This method allows to assign shares of capital and operational costs associated with imported power from generation capacities abroad, and to allocate transmission capacity costs based on the countries' role in the fluctuating spatio-temporal power flow patterns. We observe that for the model of the European system, both cooperation between the countries by importing and exporting excess generation, and a more efficient heterogeneous capacity placement not only lead to a reduction of the system LCOE, but also to a reduction of the nodal LCOE for all countries compared to the case without transmission. It is shown that net exporters are the main beneficiaries of a heterogeneous system layout, with an export flow-based attribution of transmission infrastructure cost reducing the related spread of nodal LCOEs.

The simplified model used in this study neglects various features of the electricity system which could be covered in a more exact modeling approach. The spatial scale with countries as coarse-grained nodes of the network does neither take into account the spatial distribution of generation capacity and load inside the countries, nor potential intra-country bottlenecks of the transmission system. A more realistic representation of a future highly renewable electricity system should also include generation from hydropower and the use of storage technologies like battery, hydrogen, or pumped hydro storage (see~\cite{schlachtberger2017} for an extension of the model in~\cite{eriksen2017}, which includes these technology options). Going one step further, also other energy sectors and their coupling to the electricity system have to be addressed (see for instance~\cite{Brown2018}, which follows a similar approach as~\cite{schlachtberger2017}, but for a sector-coupled energy model). Finally, both the investment and operational decisions in the electricity system are made by the system participants under the technological, economic and regulatory boundary conditions. This can be represented by dedicated market or investment models, for instance using agent based models (see~\cite{Weidlich2008} and~\cite{Ringler2016} for reviews of agent based electricity market models). For all these increasingly complex models, the system dynamics at some point lead to nodal electricity imports and exports, which in turn result in a power flow pattern on the network according to the (potentially approximated) physical power flow equations. It is this flow pattern, which is subject to the flow tracing method, irrespective of the underlying electricity system representation in the model (see also~\cite{Schaefer2017b} for a discussion of the input and output of the flow tracing method). The approach presented in this contribution thus does not only apply to the presented simplified model, but also to arbitrary models incorporating power flows on a network. 

Methodologically it would be interesting to use other flow allocation measures for the transmission cost allocation, for instance methods based on the PTDF matrix, and compare the resulting influence on the nodal LCOEs (see~\cite{Brown2015} for an exemplary application of such a method to a future scenario with high renewables penetration in Europe in 2050, and~\cite{Schaefer2017b} for a comparison of a PTDF matrix based flow allocation method with the flow tracing approach used in this contribution). Also the allocation of backup power capital costs could be expanded to include correlations to the injection and flow pattern, similar to the approach used for the network usage measure in~\cite{Tranberg2015}.

For the investigation in the present contribution we have used a coarse-grained model where the heterogeneity has been limited to country aggregates. However, the capacity factors vary within each country, which can be exploited by using a model with higher spatial resolution. Recently, in~\cite{Hoersch2016} it has shown that the cost of renewable generation capacity in a heterogeneous system is up to 10\% lower for a high resolution network of 362 nodes when compared to a 37 node network with one node per country. It would be interesting to apply the concept of nodal LCOEs and understand the distribution of costs in such a more detailed model.

The method of flow-based nodal LCOEs allocates the system costs of power generation to the loads  (consumers) in the electricity network. This approach is very different from the workings in the actual electricity system, in which the owners of generation capacity in general have to seek remuneration of their investments though transactions in different electricity markets. The pricing mechanism in these markets is usually based on the marginal cost of providing electricity or system services at a certain location, which only indirectly factors in the capital costs of generation capacity through the investment decisions of market participants. The method presented in this contribution thus first and foremost serves as an analytical tool to investigate the system costs of an interconnected electricity system on a nodal level. By definition, it does not take into account the role of markets as the prevalent way to allocate costs and remunerate investors in today's electricity system. Nevertheless, comparing a statistical measure of market prices with the nodal LCOEs could point towards how additional capacity pricing mechanisms could lead to a fair distribution of the respective costs in a cost-efficient heterogeneous highly renewable energy system (see~\cite{IEA2016} for a discussion of electricity market design and regulation for a future low-carbon power system).  Such an approach could complement the marginal-cost based principle in future electricity markets, adding a flow-based nodal allocation of certain components of system costs, or it could inspire entirely new market rules.

\section*{Acknowledgments}
Mirko Sch\"afer is funded by The Carlsberg Foundation Distinguished Postdoctoral Fellowship. Jonas H\"orsch is funded by the CoNDyNet project, which is supported by the German Federal Ministry of Education and Research under grant no. 03SF0472C. Martin Greiner is partially funded by the RE-INVEST project (Renewable Energy Investment Strategies -- A two-dimensional interconnectivity approach), which is supported by Innovation Fund Denmark (6154-00022B). The responsibility for the contents lies solely with the authors.

\section*{Bibliography}
\bibliographystyle{unsrt}
\bibliography{references.bib}
\end{document}